\documentclass[a4paper,twocolumn,prl,superscriptaddress]{revtex4}
\usepackage{amsmath,amssymb}
\usepackage{graphicx,epstopdf}
\usepackage{color}
\usepackage{subfigure}
\usepackage{hyperref}

\begin{document}
\title{Q-Gaussian diffusion in stock markets}
\author{Fernando Alonso-Marroquin}
\affiliation{School of Civil Engineering, The University of Sydney, Australia}
\email{fernando.alonso@sydney.edu.au }
\author{Karina Arias-Calluari}
\affiliation{School of Civil Engineering, The University of Sydney, Australia}
\author{Michael Harr\'e}
\affiliation{School of Civil Engineering, The University of Sydney, Australia}
\author{Morteza. N. Najafi}
\affiliation{Department of Physics, University of Mohaghegh Ardabili, Ardabil, Iran}
\author{Hans J. Herrmann}
\affiliation{PMMH, ESPCI, 7 Quai St. Bernard, 75005 Paris, France}

\begin{abstract}
We analyze the Standard \& Poor's 500 stock market index from the last $22$ years. The probability density function of price returns exhibits two well-distinguished regimes with self-similar structure:
the first one displays strong super-diffusion together with short-time correlations, and the second one corresponds to weak super-diffusion with weak time correlations. Both regimes are well-described by q-Gaussian distributions. The porous media equation is used to derive the governing equation for these regimes, and the Black-Scholes diffusion coefficient is explicitly obtained from the governing equation. 
\end{abstract}
\maketitle

Price fluctuations in stock markets exhibit remarkable features such as strong short-time correlations, weak long-time correlations, power law tails, and slow convergence to the normal distribution \cite{vasconcelos2004guided}. In the earliest stock market model, Bachelier proposed classical Brownian motion to represent price fluctuations. This model was the cornerstone for the well-established Black-Scholes equations for stock markets. Mandelbrot suggested that classical diffusion was not appropriate for modeling real stock markets \cite{mandelbrot1963variation}. His conclusion was based on the analysis of price variations of the cotton index whose probability density function (pdf) was better described by a L\'evy distribution. Later, Mantegna and Stanley proposed that this L\'evy distribution should be truncated to achieve consistency with the slow convergence to normality, and to guarantee that the standard deviation of price variations remains finite \cite{mantegna1994stochastic}. Most recent developments have suggested that the q-Gaussian distribution ---an extension of the Gaussian distribution for correlated fluctuations--- is more appropriate to account for the correlations of price increments in the NASDAQ stock market index \cite{borland2002option} and NYSE index \cite{tsallis2003nonextensive}. Time correlations lead to anomalous diffusion, a phenomenon that is pervasive in strongly correlated classical systems, such as silo discharge \cite{arevalo2007anomalous} and sheared granular flow \cite{combe2015experimental}.

In this Letter we have analyzed the Standard \& Poor's $500$ (S\&P500) stock market data during the $22$-year period from January $1996$ to May $2018$, with an interval span of $1$ min. The stock market index at time $t$ is denoted by $I(t)$. The price return in a time interval from $t_0$ to $t$ is defined by
\begin{equation}\label{eq:Price return}
X(t,t_0)=I(t_0+ t)-I(t_0).
\end{equation}
The stock market index fluctuates over time in a random fashion. The main interest of economists is to predict the price return at any future time $t_0+t$. Here we adopt the probabilistic approach: we assume that the price return X is a random variable with  probability density function (pdf) $P_X (x,t)$. Then we formulate the governing equation for this distribution. The standard diffusion process is an oversimplification, as the price fluctuations strongly correlate for times of the order of minutes, and they weakly correlate for longer times \cite{mantegna1999introduction}. A candidate for such a distribution function is the q-Gaussian distribution. The q-Gaussian is a generalization of the Gaussian distribution, and is defined as \cite{umarov2008q}:
\begin{equation}\label{eq:q-Gaussian}
g_q(x,\beta)=\frac{\sqrt{\beta}}{C_{q}}e_{q}(-\beta x^{2}),
\end{equation}
where $e_{q}(x)=\left[ 1+(1-q)x\right]^\frac{1}{1-q} $ is the q-exponential function. For $q>1$, the q-Gaussian has asymptotic heavy-tail power law given by $g_q\sim1/x^{2/(q-1)}$. The ``q-Gauss" is a special case of q-Gaussian distribution defined as $g_q(x)=g_q(x,\beta=1))$. The exponential and Gaussian functions can be recovered by taking the limit $q\to1$. For $1<q<3$, the normalizing constant $C_q$ is given by:
\begin{equation}\label{eq:Cq}
C_{q}=\sqrt{\frac{\pi}{q-1}}\frac{\Gamma(\frac{3-q}{2(q-1)})} {\Gamma( \frac{1}{q-1}) }.
\end{equation}

\begin{figure*} [tb]
\centering
\includegraphics[scale=0.32,trim=0.0cm 0cm 0cm 0cm,clip=true]{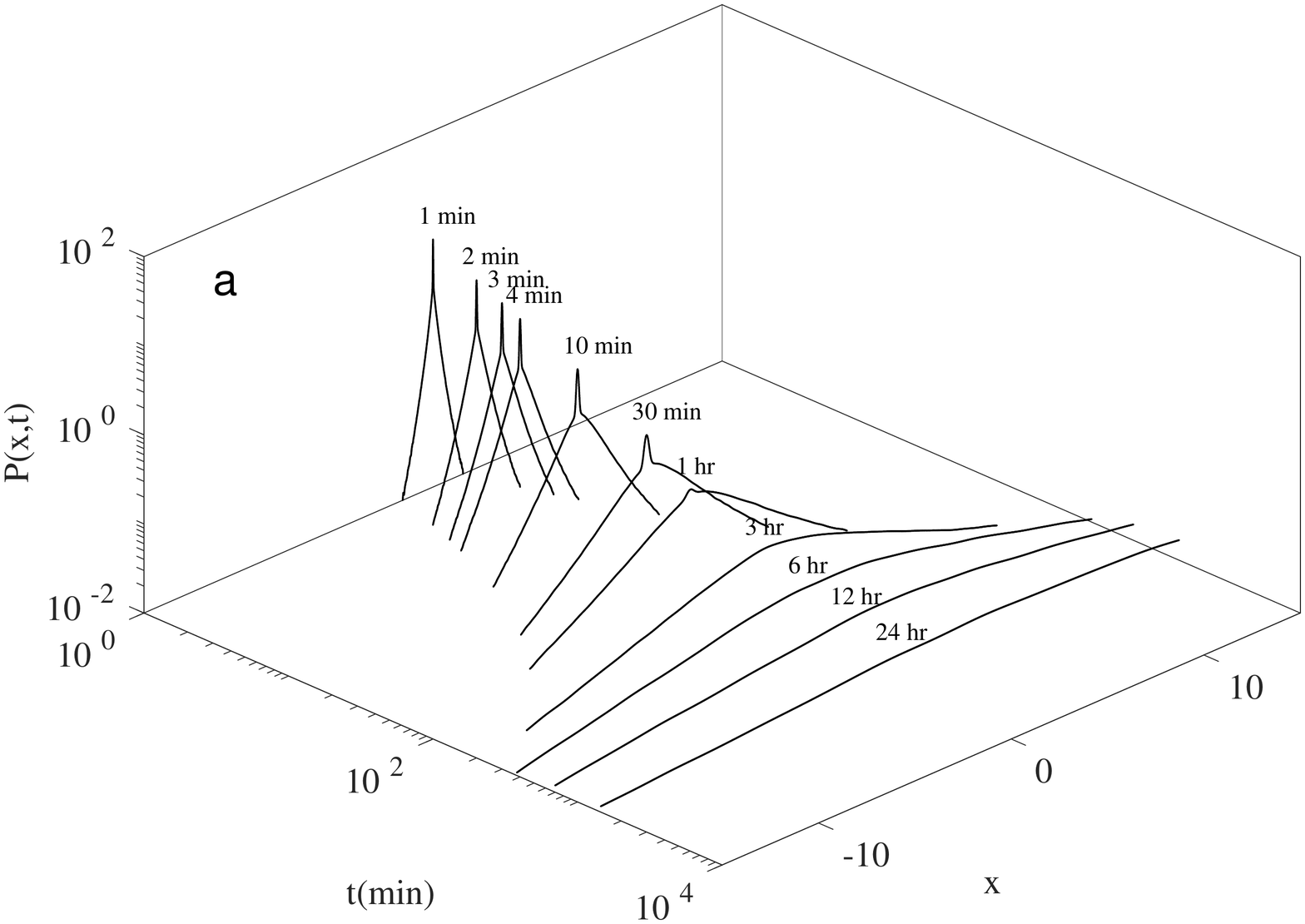}
\includegraphics[scale=0.28,trim=0.0cm 0cm 0cm 0cm,clip=true]{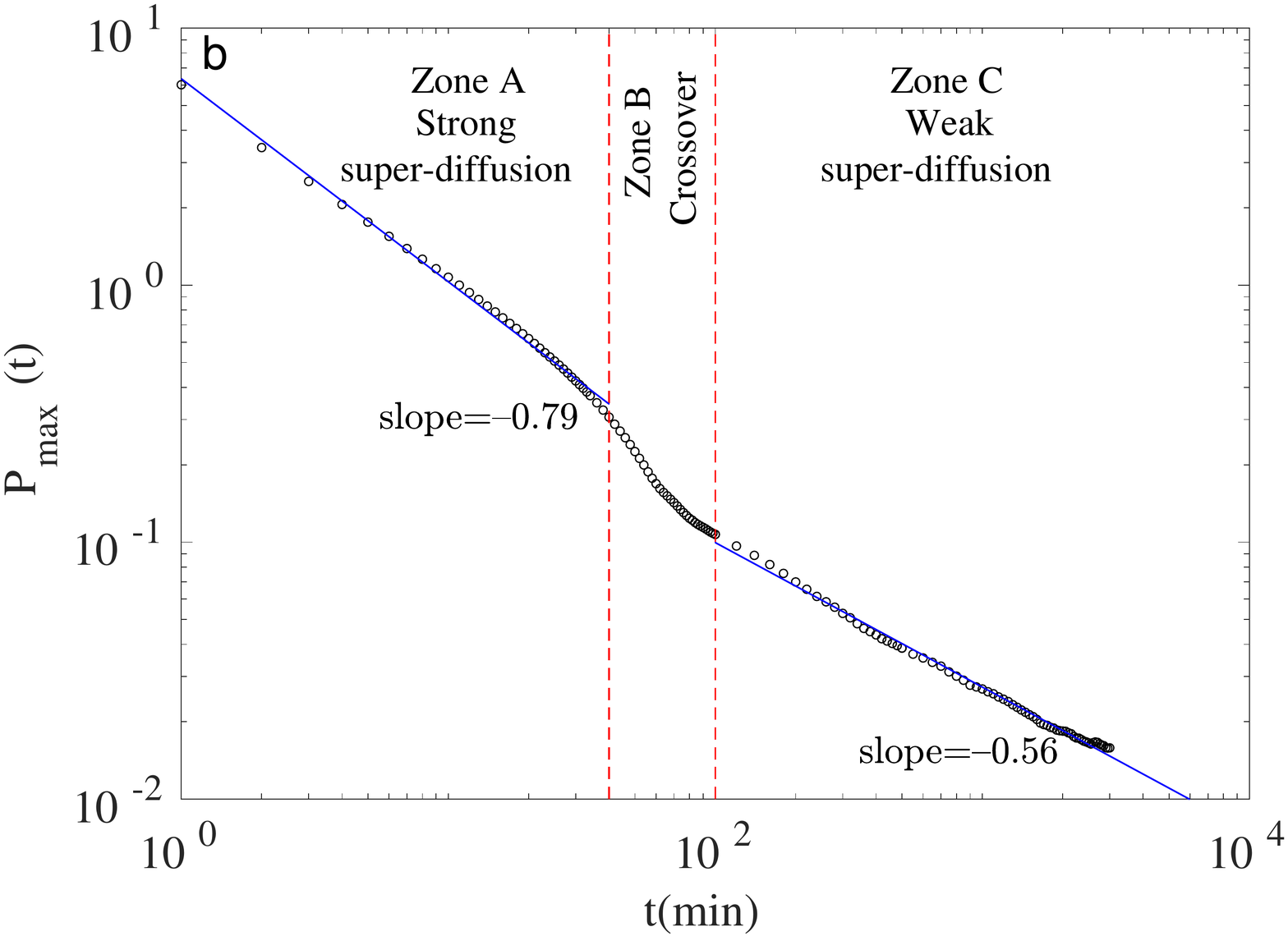}
\includegraphics[scale=0.25,trim=0.0cm 0cm 0cm 9cm,clip=true]{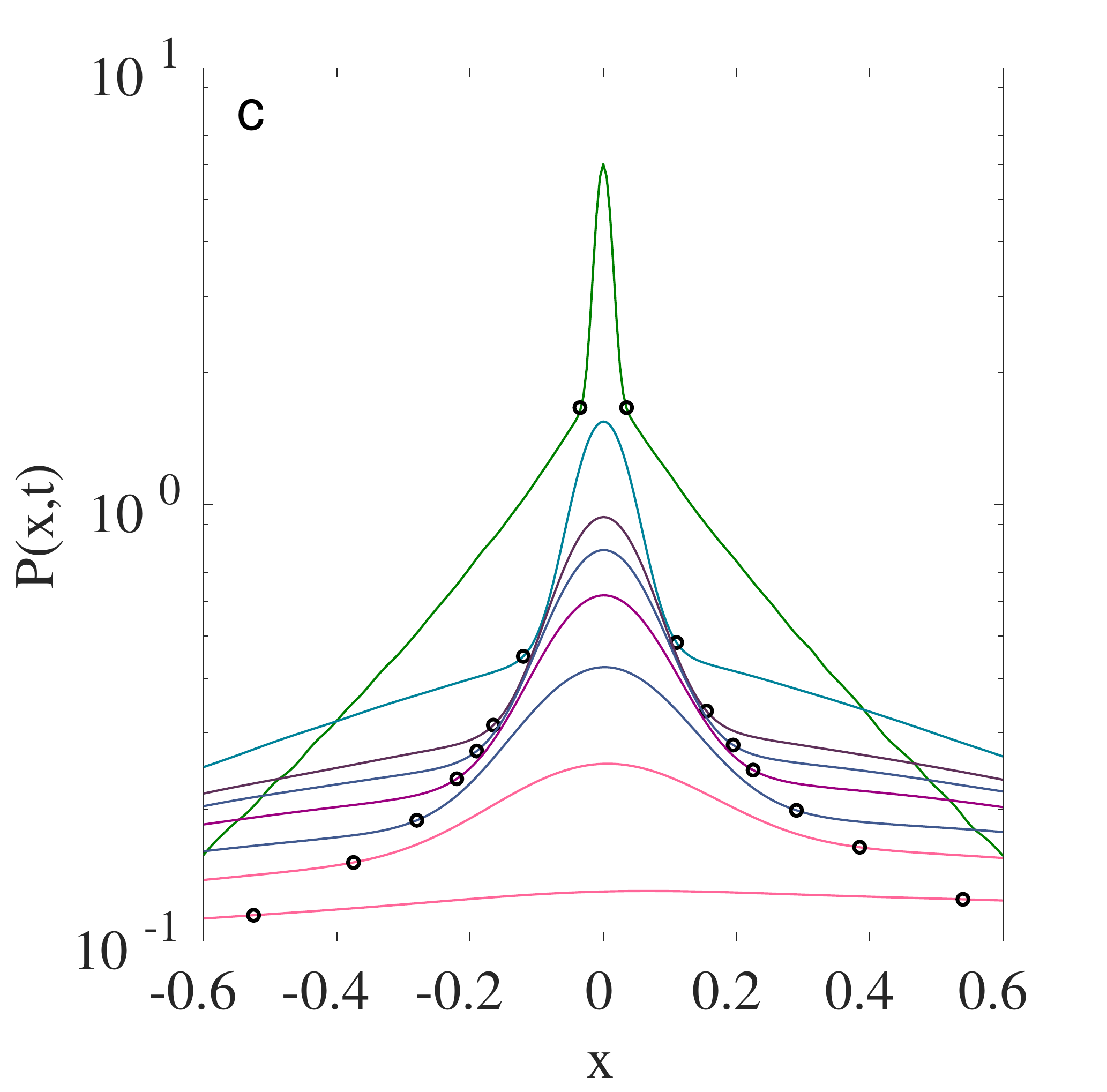}
\includegraphics[scale=0.25,trim=0.0cm 0cm 0cm 9cm,clip=true]{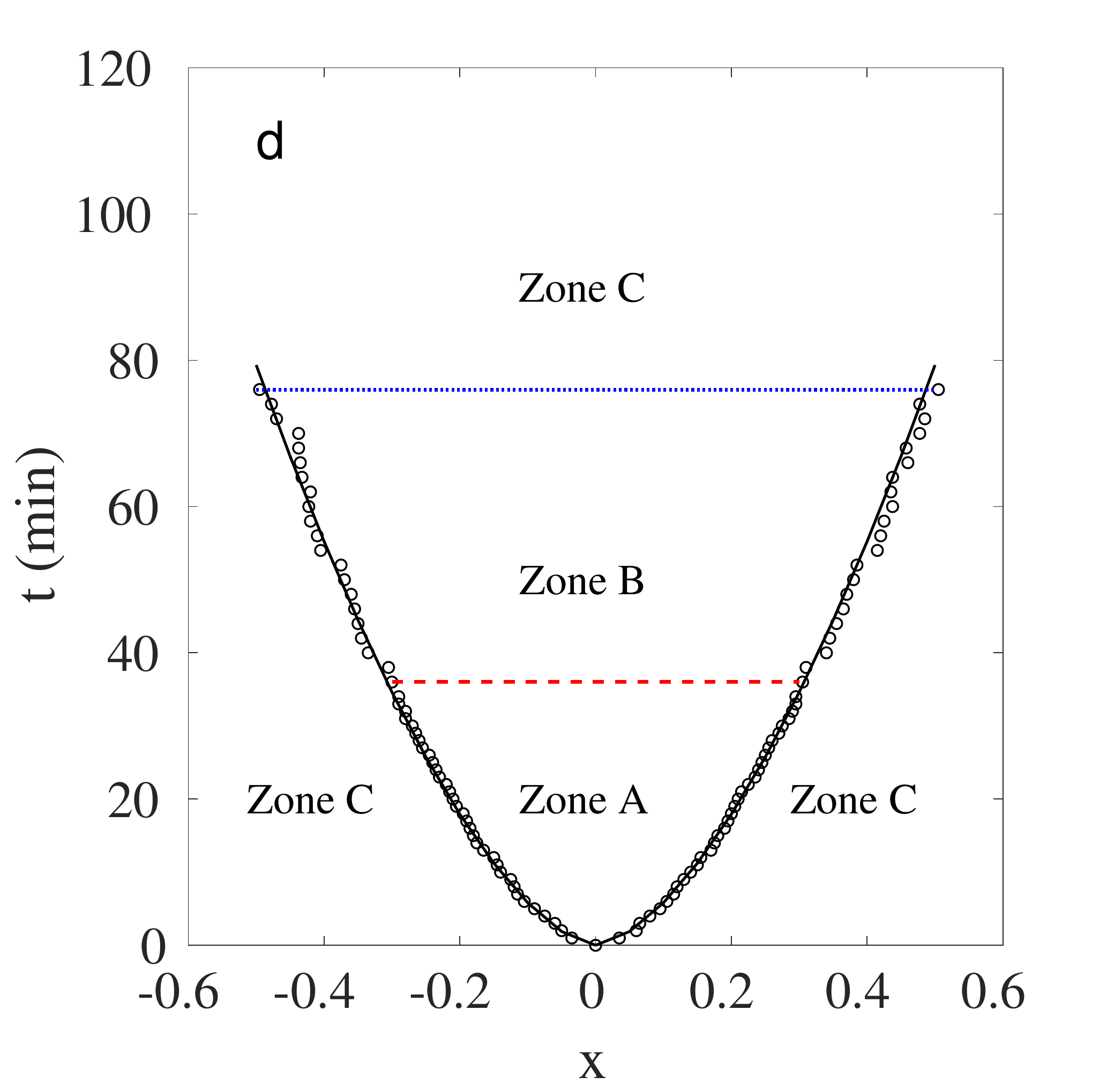}
\includegraphics[scale=0.25,trim=0.0cm 0cm 0cm 9cm,clip=true]{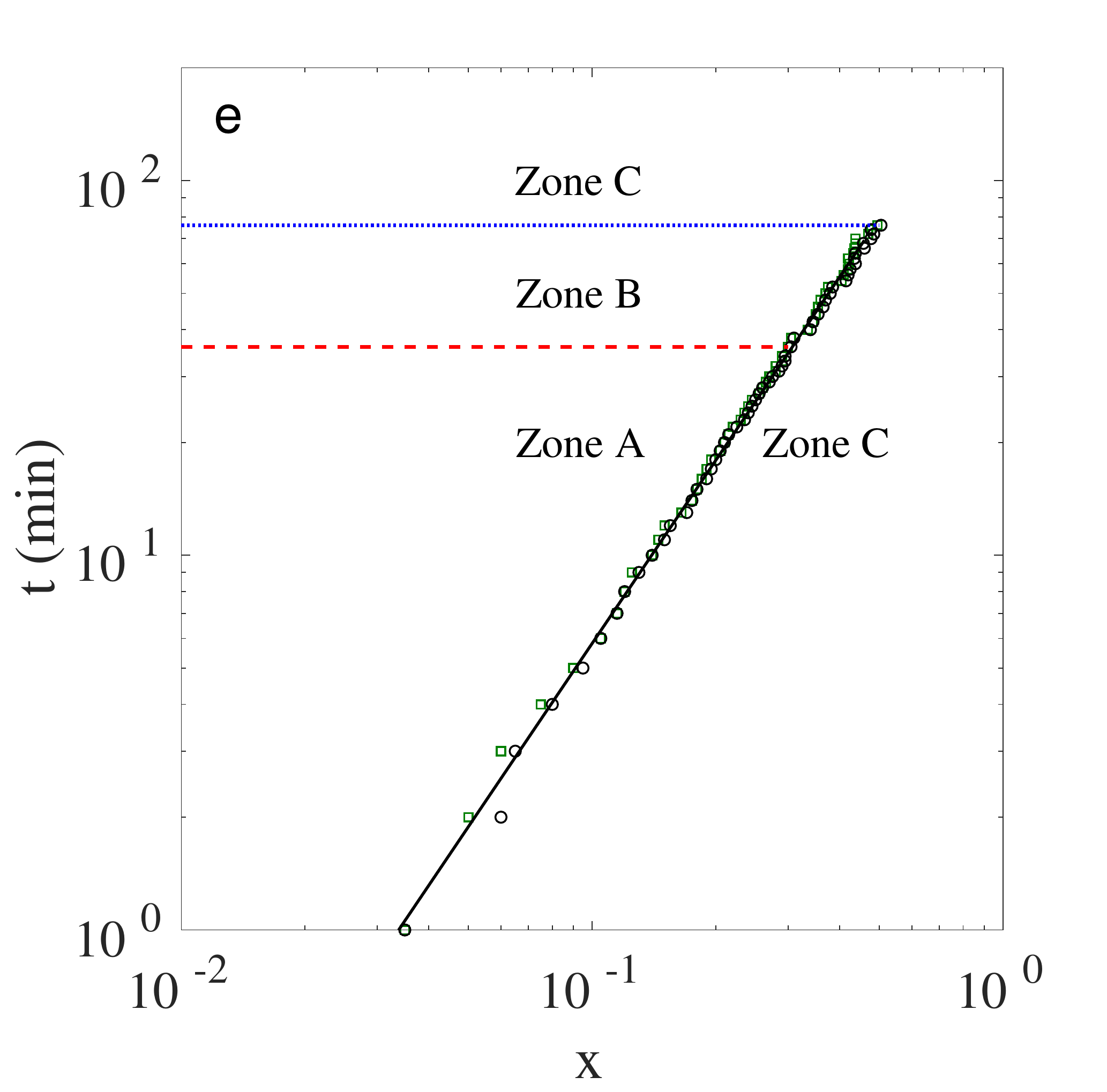}
\caption{(Color online) 
(a) Time evolution of the pdf of price return. Initially the pdf has a pronounced bump in the center that fully disappears close to $78$ minutes.  (b) The time evolution of the height of the pdf. Two well-defined power laws are observed. (c) From Figure \ref{fig:regimes}a, time increases from top to bottom. The ends of the bump is obtained from the two point at the pdf with abrupt change of slope. These points correspond to a transition from strong to weak super-diffusion. (d)The circles represent the end points plotted against time that are fitted in (e) by the power law $x=\pm a (t/t_o)^\nu$, with $a= (3.39 \pm 0.01)\times 10^{-2}$, $t_o=1$ min and $\nu=0.62\pm0.05$. This curve and the line $t=35$ min  (red dotted line) define the strong super-diffusion regime (zone A). The bump disappears completely at $t = 78$ mins (blue dashed line). The remaining area corresponds to the weak super-diffusion regime (zone C). The crossover regime (zone B) is limited by the curve and $35$ mins $<t< 78$ mins. In this regime the bump still has not dissipated but experiences a transition from strong to weak super-diffusion.}
\label{fig:regimes}
\end{figure*}

The so-called q-Central Limit Theorem states that the q-Gaussian is the limit of the distributions of specially correlated random processes \cite{umarov2008q}. Thus, it is reasonable to propose the q-Gaussian as a candidate to fit the pdf distribution of stock markets. With this aim, we construct the pdf of the S\&P500 stock market index using the kernel density estimator. The bandwidth of the kernel is set to $h= 0.005$ that is small enough to capture the non-trivial structure of the pdfs. Figures \ref{fig:regimes}a and \ref{fig:regimes}b show the time evolution of the pdf and its height from $1$ minute to $24$ hours of active market time. 
The initial distribution at $t=1$ minute consists of heavy tails and a pronounced bump at the center. This bump is easily distinguished from the rest of the distribution by an abrupt change of the slope of the distribution, see Figure~\ref{fig:regimes}c.  The points where the abrupt change of the slope occurs are plotted against the time in Figures~\ref{fig:regimes}d and ~\ref{fig:regimes}e. These points define the top and bottom boundary of what we called the domain of the bump. As  time evolves, the bump diffuses and completely disappears after $78$ minutes.


As shown in Figure \ref{fig:regimes}b, the time variation of the height of the bump obeys a power law with exponent $P_{max}\sim t^{-1/\alpha}$ with $\alpha =1.26 \pm 0.04$ in the strong super-diffusion regime. This is different from the exponent $\alpha=2$ expected in classical diffusion processes. Between $t=38$ mins and $t=78$ mins, we observe a crossover region. The end of the crossover corresponds to the region where the bump fully disappears, which is shown in Fig \ref{fig:regimes}d and \ref{fig:regimes}e. After the end of the crossover, the new height of the distribution obeys a different power law with exponent $\alpha = 1.79 \pm 0.01$  which is closer to the exponent of classical diffusion. Mantegna and Stanley's analysis on a more limited data-set of the S\&P500 index led to the exponent $\alpha=1.40 \pm 0.05$ \cite{mantegna1995scaling}. This is in reasonable agreement with our exponents, considering that they used a single exponent to fit the entire time evolution of the height. 

Based on the domain of the bump and the time evolution of the height of the pdf, we partition the two-dimensional space (price and time) into three zones as shown in the Figure \ref{fig:regimes}d:  Zone A is the domain of the bump where the power law holds, zone B is the area of the bump's domain where the power law smoothly changes to another power law, and zone C is the remaining space. In zones A and C we propose a self-similar distribution given by
\begin{equation}\label{eq:self-similar}
P(x,t)= \frac{1}{(Dt)^\frac{1}{\alpha} }f\left( \frac{x}{(Dt)^\frac{1}{\alpha}}\right).
\end{equation}
Where $f(x)$ is a normalized distribution. The exponent $\alpha$ defines the diffusion process as follows: The second moment of the distribution in Eq.~\ref{eq:self-similar}  is $\langle x^2\rangle\sim t^{2/\alpha}$. Then $\alpha <2$ corresponds to super-diffusion, whereas $\alpha > 2$ leads to sub-diffusion. The exponent $\alpha$ scales the height of the distribution as $P_{max}\sim t^{-1/\alpha}$ as the previous fitting in Figure \ref{fig:regimes}b. The present favorite alternative is $f(x)=L_\alpha (x)$ the L\'evy distribution, as proposed by Mandelbrot \cite{mandelbrot1963variation}, and Mantegna and Stanley \cite{mantegna1995scaling}. They obtained $\alpha = 1.4$ as a fitted parameter, indicating super-diffusion.  Here we propose a different approach by seeking a self-similar fitting for both weak and strong super-diffusion regime using the q-Gauss function $f(x)=g_q(x)$ in Eq.~\ref{eq:self-similar}. In both models the classical diffusion can be recovered by taking $\alpha=2$ and $g(x)=g_1(x)= L_2(x)$. This limit corresponds to the self-similar solution of the diffusion equation that does not fit well to the stock market data.

\begin{figure}[t]
	\begin{center}
		\includegraphics[width=8 cm, trim=0.0cm 0.2cm 1cm 1.0cm,clip=true]{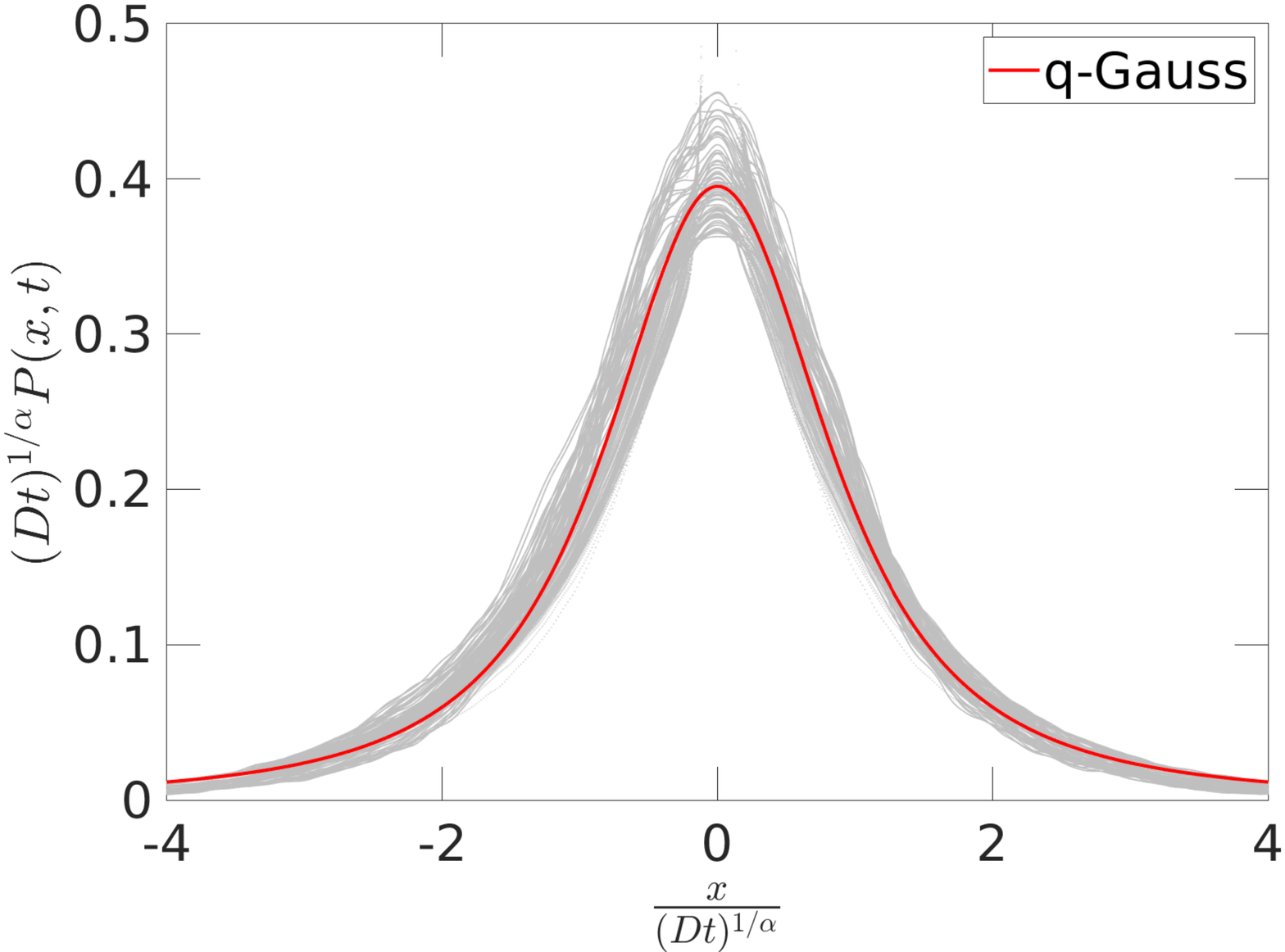}
	\end{center}
	\caption{(Color online) Collapse of the pdfs of the weak super-diffusion regime (zone C). The exponent used for the collapse is   $\alpha = 1.79 \pm 0.01$. The collapse data is fitted by a q-Gauss (red line) with exponent $q=1.71\pm0.01$ and $D=0.1118 \pm 0.005$ min$^{-1}$ . Both exponents are related by the Eq.~\ref{eq:alpha-q}.  }
	\label{fig:weak}
\end{figure}

\begin{figure}[t]
	\begin{center}
		\includegraphics[width=8 cm, trim=0.0cm 0.0cm 1cm 1cm,clip=true]{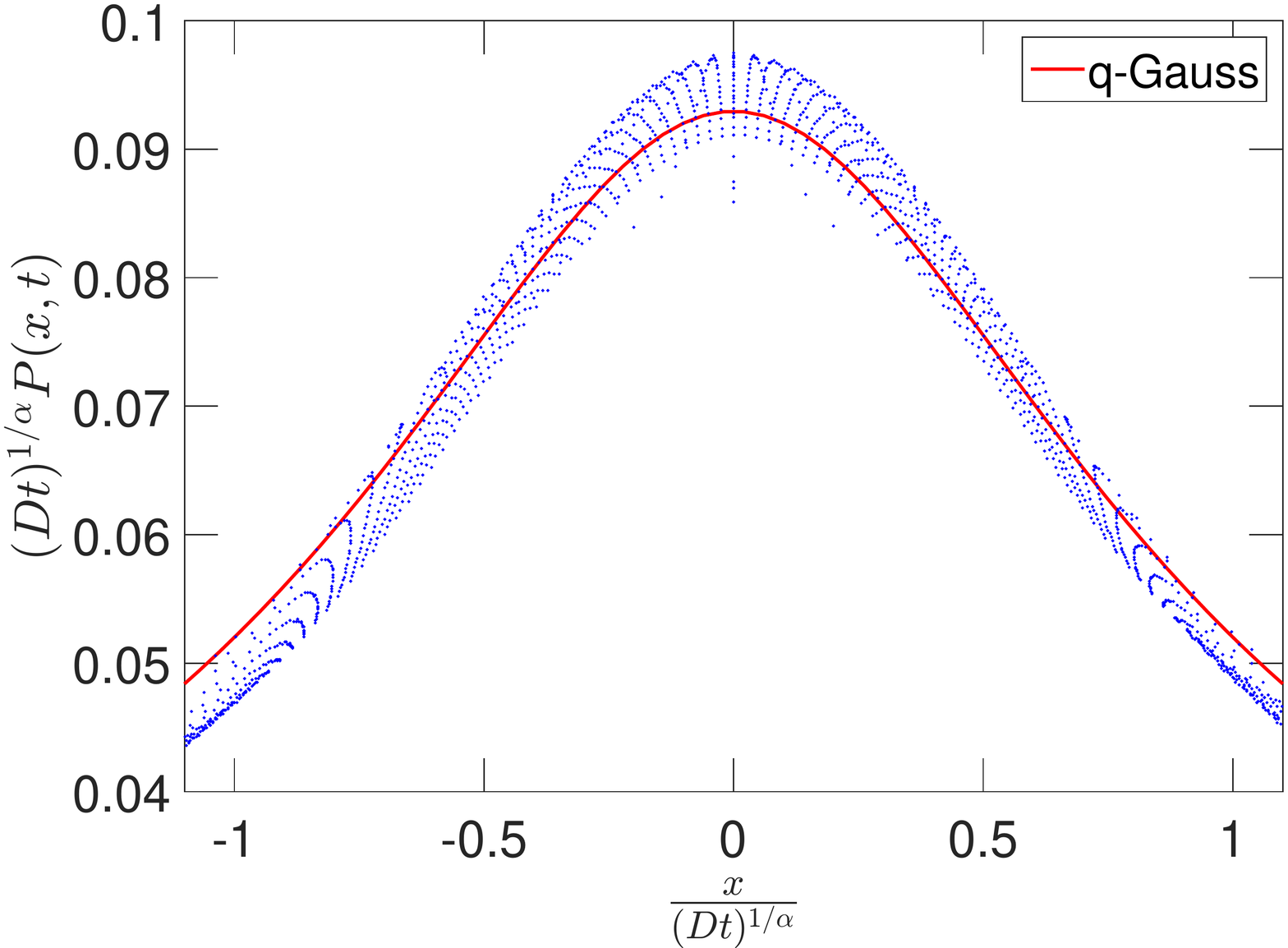}
	\end{center}
	\caption{(Color online) Collapse of the pdfs of the strong super-diffusion regime (zone A). The exponent used for the collapse is $\alpha = 1.26 \pm 0.04$ and 
		$D=(4.8\pm 0.2)\times 10^{-3}$ min$^{-1}$. The collapse data is fitted by a q-Gauss (red line) with exponent $q=2.73 \pm 0.005$.}
	\label{fig:strong}
\end{figure}

The self-similar fitting in zones A and C is performed as follows: First we fit each pdf to Eq.~\ref{eq:q-Gaussian} using $q$ and $\beta$ as fitted parameters. We evaluate the time dependence of the fitting parameters. For each zone, we found that $q$ is approximately constant while $\beta$ follows a power law relation that is written as $\beta = (Dt)^{-2/\alpha}$, where $D$ and $\alpha$ are fitting parameters of this power law. Then, we collapse the pdfs for both weak and strong super-diffusion, as shown in Figures \ref{fig:weak} and \ref{fig:strong}.

To collapse the pdfs in the weak super-diffusion regime ---zone C in Figure \ref{fig:regimes}d---  we use the data from $t = 1$ minute to $t = 3000$ minutes. The data is detrended by subtracting from the time series the average value within a time window of  one month. This removes the effect of the drift on the pdf. We obtain an excellent agreement for the collapsed data with the q-Gauss distribution (Eq.~\ref{eq:q-Gaussian} with $\beta =1$). The q-exponent in this regime is $q=1.72 \pm 0.03$ which is larger than the value $q=1$ expected for uncorrelated random processes.  This is consistent with the weak correlation of the price fluctuations in this regime that is given by an auto-correlation of price fluctuating around  $ 0.1 \%$. The exponent $\alpha = 1.79 \pm 0.01$ is the same as the one calculated in Figure \ref{fig:regimes}b and is lower than the value of $2$ expected for classic diffusion.
The collapse of the pdfs in the strong super-diffusion regime ---zone A in Figure \ref{fig:regimes}d--- is shown in Figure~\ref{fig:strong} Like in the weak super-diffusion case, the collapse fits well to the q-Gauss distribution. However, in this case the exponent $q=2.73 \pm 0.005$ is larger and $\alpha = 1.26 \pm 0.04$ is lower than the exponents in the week super-diffusion regime. This indicates stronger deviation from classical diffusion. As in the previous case, these exponents should be considered as independent exponents. Note that in Figure~\ref{fig:strong}  we collapse only the bump of the distribution, while its tail is fitted in Figure~\ref{fig:weak} because it belongs to the weak super-difussion regime (zone C).

Next we construct the evolution equation of the pdf based on the q-Gaussian fitting. It is natural to relate the super-diffusion processes with a well-known anomalous diffusion model, which is given by a non-linear Fokker-Planck equation  \cite{borland2002option}, also known as the {\it porous media equation} \cite{tsallis2009introduction}:
\begin{equation}\label{eq:PME}
\frac{\partial u}{\partial t}=\frac{\partial ^{2}u^{m}}{\partial x ^{2} } .
\end{equation}
The Barenblatt solution of Eq. \ref{eq:PME} for $m<2$ and $t>0$ is given by  \cite{vasquez2007porus} 
\begin{equation}\label{eq:BarenblattA}
u_m(x,t)=\frac{1}{t^\frac{1}{m+1} }\left(  C + \frac{m-1}{2m(m+1)}\frac{x^2}{t^\frac{2}{m+1}}  \right)^\frac{1}{m-1}. 
\end{equation}

Where C is an integration constant. There is only one exponent in Eq. \ref{eq:PME} while we have two independent exponents $\alpha$ and $q$ in the fitting of the collapsed data (see Figures   \ref{fig:weak} and \ref{fig:strong}). An additional exponent $\xi$ is introduced and proposing $m=2-q$ we write
\begin{equation}\label{eq:P-u}
P(x,t) = u_{2-q}(x,\tau) \quad \quad \tau=(Bt)^\xi. 
\end{equation}
The scale parameter $B$ is obtained as follows: By placing Eq. \ref{eq:P-u} into Eq. \ref{eq:BarenblattA} we obtain,  
\begin{equation}
	P(x,t)=\frac{1}{(Bt)^\frac{\xi}{3-q} }\left(  C - \frac{1-q}{2(2-q)(3-q)}\frac{x^2}{(Bt)^\frac{2\xi}{3-q}}  \right)^\frac{1}{1-q}.\nonumber
\end{equation}
This equation can be writen as:
\begin{equation}
P(x,t)= \frac{1}{C_q(D t)^\frac{1}{\alpha} }\left(  1 - (1-q)\frac{x^2}{(D t)^\frac{2}{\alpha}}  \right)^\frac{1}{1-q},
\label{eq:P-u_2}
\end{equation}
in which $C_q$ is the normalization constant in Eq.~\ref{eq:Cq}. The $B$ and $C$ values are related to $D$ and $C_q$ by

\begin{equation}\label{eq:C-q}
\begin{split}
&C_{q}={B^{{\frac{1}{\alpha}}}}C^{\frac{1}{q-1}}D^{-\frac{1}{\alpha}}\\
&D=B(2C(2-q)(3-q))^{\alpha/2} 
\end{split}
\end{equation}
and $\xi$ is calculated in terms of $\alpha$ and $q$ by 
\begin{equation}\label{eq:alpha-q}
\xi =\frac{3-q}{\alpha}.
\end{equation}
The parameters $\alpha$, $q$ and $D$ are derived by fitting the collapse data. Eq.~\ref{eq:P-u_2} corresponds to Eq.~\ref{eq:self-similar} with $f(x) = g_q(x)$ defined by Eq.~\ref{eq:q-Gaussian}.  Finally, the governing equation of $P(x,t)$ is obtained by first taking the partial time derivative in Eq.~\ref{eq:P-u} 
\begin{equation}\label{eq:chain rule}
\frac{\partial P}{\partial t} =\frac{\partial P} {\partial \tau} \frac{\partial \tau} {\partial t} = \frac{\partial u_{2-q}}{\partial t} (Bt)^{\xi-1} \xi B. 
\end{equation}
Placing Eq.~\ref{eq:chain rule} into Eq. \ref{eq:PME} the governing equation is obtained
\begin{equation}\label{eq:NLFP}
t^{1-\xi}\frac{\partial P}{\partial t}=\xi D^\xi \frac{\partial ^{2}P^{2-q}}{\partial x ^{2} }.
\end{equation}

\begin{figure}[htp]
\begin{center}
\includegraphics[width=8 cm, trim=1.0cm 2.0cm 1cm 2cm]{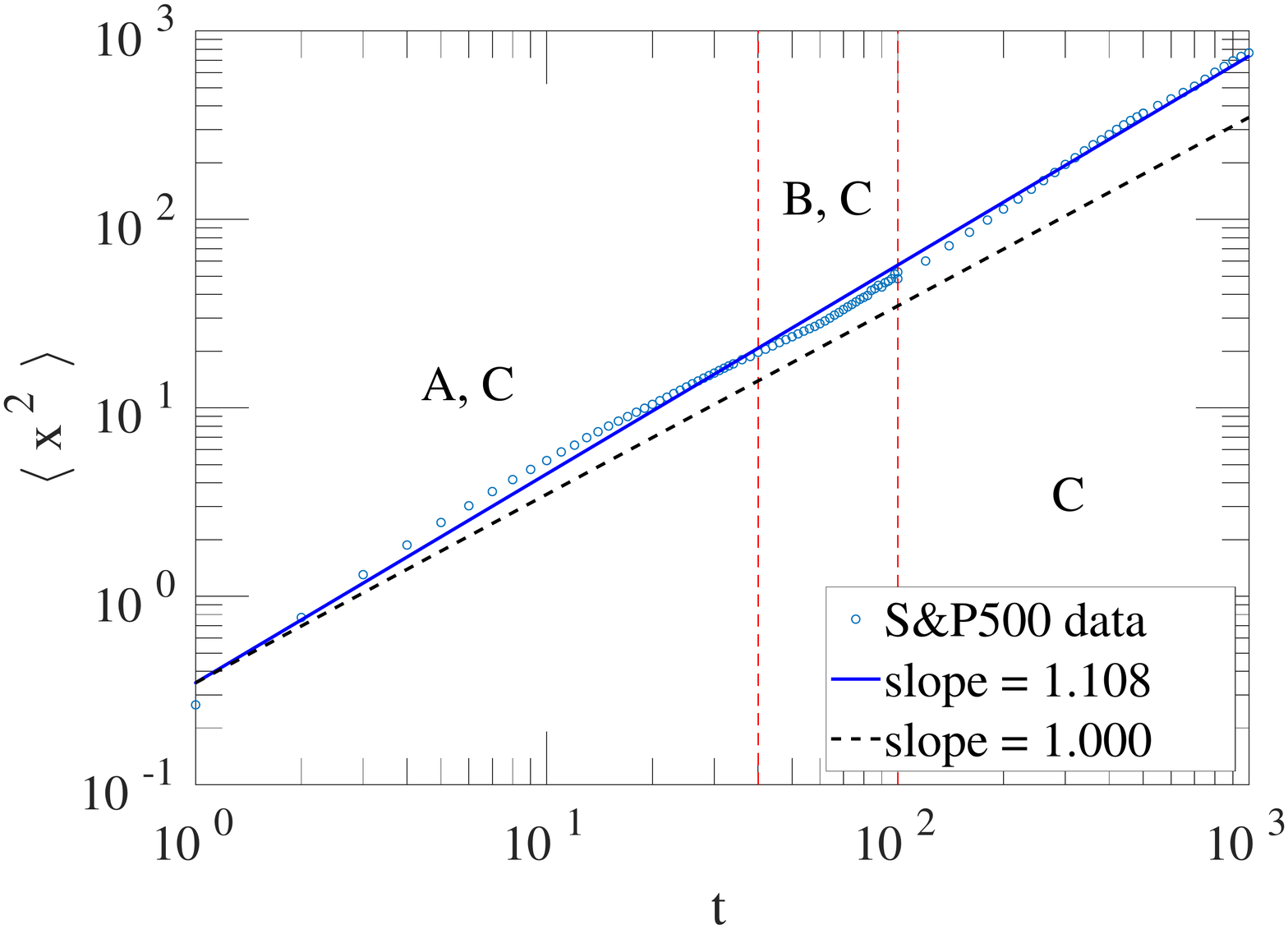}
\end{center}
\caption{(Color online) Time evolution of the second moment of the full pdf of the S\&P500 index. The data is compared to the best fitting and the linear fitting that corresponds to classical diffusion. The zones where the integration is performed are also shown.}
\label{fig:moment}
\end{figure}

Eq. \ref{eq:NLFP} is of great importance as it is the evolution equation of the distribution of price returns of the S\&P500 index. The parameters $q$, $\xi$ and $D$ take different values depending on whether we are in the weak or strong super-diffusion regime. Eq. \ref{eq:NLFP} reflects a distinct feature of stock markets, namely that the diffusion coefficient depends on the diffused quantity itself, a feature that is observed also in many biological and physical systems \cite{Bogachev2015FPE,Till2005nonlinear}. Also, the Black-Scholes coefficient of diffusion can be calculated by comparing Eq.~\ref{eq:NLFP}  to the  linear Fokker-Planck equation $\frac{\partial P}{\partial t} =\frac{\partial^2(D_2 P)} {\partial x^2}$ \cite{vasconcelos2004guided}. The direct comparison leads to $D_2=  \xi D^\xi P^{1-q}t^{\xi-1}$. Replacing Eqs.~\ref{eq:P-u_2} and ~\ref{eq:alpha-q} into this equation we obtain an explicit expression for the coefficient of diffusion:
\begin{equation}\label{eq:difussion}
D_2(x,t) = \frac{(3-q) D^\frac{2}{\alpha} } {\alpha C_q^{1-q} t^\frac{\alpha-2}{\alpha} } \left(  1 - (1-q)\frac{x^2}{(D t)^\frac{2}{\alpha}}\right).
\end{equation}
For a fixed time and large price fluctuations, the scaling $D_2\sim x^2$ of the Black-Scholes equation for geometric Brownian motion is recovered \cite{vasconcelos2004guided}. Our extension introduces the power law dependency with time into this relation that accounts for super-diffusion.

The strong super-diffusion regime occurs in a very small zone in the phase space, so that it only contributes to the moments of the pdf distributions to a small extent. We characterize the {\it global diffusion} by plotting the second moment of the full pdf as a function of time, and fitting it to the best power law. The result is shown in Figure ~\ref{fig:moment}. The second moment is almost linear with time, suggesting that the global diffusion is weakly super-diffusive.

We have identified two regimes in the time evolution of the pdfs of the price return of the S\&P500 index that account for strong and weak super-diffusion. Both regimes are described by the family of self-similar q-Gaussian distributions that accurately capture the tail distributions, needed for financial risk estimates, by accounting for the strong super-diffusion in the centre of the distribution. The time evolution of the pdf is consistent with the central limit theorem for correlated fluctuations. The strong correlations in the short-time regime is reflected in a strong super-diffusive q-Gaussian regime that has not previously been reported, and the weak correlation in the long-time regime corresponds to the weak super-diffusive q-Gaussian regime. We demonstrated that these regimes are well-described by a non-linear Fokker Plank equation that is used to obtain an explicit formula for the Black-Scholes diffusion coefficient.

HJH thanks CAPES and FUNCAP for support. MH acknowledges Australian Research Council grant DP170102927.
\bibliographystyle{prsty}
\bibliography{stockmarkets}

\end{document}